# The Weaves Reconfigurable Programming Framework

Srinidhi Varadarajan
Department of Computer Science
Virginia Tech, Blacksburg, VA 24061-0106
(srinidhi@cs.vt.edu)

## 1  Abstract

This research proposes a language independent intra-process framework for object based composition of unmodified code modules. Intuitively, the two major programming models - threads and processes - can be considered as extremes along a sharing axis. Multiple threads through a process share all global state, whereas instances of a process (or independent processes) share no global state. **Weaves** provide the generalized framework that allows arbitrary (selective) sharing of state between multiple control flows through a process. In the Weaves framework a single process has the same level of complexity as a workstation, with independent "sub-processes", state sharing and scheduling, all of which is achieved without requiring any modification to existing code bases. Furthermore, the framework allows dynamic instantiation of code modules and control flows through them.

In effect, weaves create intra-process modules (similar to objects in OOP) from code written in any language. Applications can be built by instantiating Weaves to form **Tapestries** of dynamically interacting code. The Weaves paradigm allows objects to be arbitrarily shared – it is a true superset of both processes as well as threads, with code sharing and fast context switching time similar to threads. Weaves do not require any special support from either the language or application code - practically any code can be weaved. Weaves also include support for *fast automatic checkpointing and recovery with no application support*. This paper presents the elements of the Weaves framework and results from our implementation that works by reverse-analyzing source-code independent ELF object files. The current implementation has been validated over Sweep3D, a benchmark for 3D discrete ordinates neutron transport [Koch et al., 1992], and a user-level port of the Linux 2.4 family kernel TCP/IP protocol stack. Performance results show that the context switch overhead in the Weaves framework is almost identical to threads.

## 2  Introduction

With the growing complexity of software code bases, application designers are increasingly moving towards component based models, where applications are built by compositional modeling. The success of the object oriented programming paradigm, with its support for encapsulation and compartmentalization of functionality, has played a major role in facilitating this mode of application development. However, to make effective use of the OO paradigm, codes need to be written in a language that supports OOP. While this is the approach of choice for the next generation of applications, we argue that it is infeasible for several areas. Areas such as scientific computing have vast repositories of legacy non-OO codes. Rewriting and validating these applications in an OO language is an enormous software engineering endeavor.

Furthermore, with the development of increasingly sophisticated sensors, compute devices have begun interacting with the real world. This interaction presents a new dimension to application adaptivity. Given the vast stream of data that flows from sensor arrays, it may not be possible to apriori determine all the possible modes in which an application may interact with the environment.  In this scenario, applications would need to evolve both in code and state to meet the needs of a dynamically changing environment, a notion we call **reconfigurable programming**.

This proposal presents a new *language independent* intra-process framework called Weaves, for object based composition of unmodified code modules. In effect, weaves create intra-process modules (similar to objects in OOP) from code written in any language. Applications can be built by instantiating Weaves to form Tapestries of dynamically interacting code. The Weaves paradigm allows objects to be arbitrarily shared - it is a true superset of both processes as well as threads, with code sharing and fast context switching time similar to threads. The main feature of Weaves is that it does not require any special support from either the language or application code - practically any code can be weaved. Our current prototype implementation uses post-compiler analysis on ELF object files, supporting the vast majority of UNIX platforms and language compilers that use ELF format.

The Weaves framework being very general, presents several new directions for developing high-performance codes. First, it enables run time composition of legacy applications, without requiring that any code be rewritten or even modified. Second, weaves enables run-time adaptivity. Run-time adaptation requires support for checkpointing and recovery and support for dynamic insertion and removal of code modules. Currently, the majority of applications implement their own user-level checkpointing, which adds a lot to the complexity and maintainability of these code bases. The Weaves framework includes transparent support for fast checkpointing and recovery with no application support. Our post-compiler analysis automatically determines the necessary state that needs to be saved and restored and presents a simple interface to this functionality. Finally, weaves support run-time flow migration, which enables automatic load-balancing of composed codes on parallel machines. The ability to migrate parts of an application as opposed to the whole gives finer grain control over load-balancing decisions.

This paper is organized as follows. In section 2 we motivate the needs for a transparent component-based framework. Section 3 describes the elements of the Weaves framework, including support for tuple spaces, automatic checkpointing and recovery, and code migration. In section 4, we present implementation details and evaluation results from the current prototype. Section 5 presents ongoing work in categorizing adaptivity, augmenting our declarative tapestry configuration with a high-level functional composition specification language, and using Weaves to provide run-time systems support for migration of parallel communication codes in a computational grid environment.

## 3  Motivation

The primary motivating application for the Weaves framework is a large-scale network emulation test-bed called the Open Network Emulator (**ONE**). The goals of the ONE project are

1. To provide a protocol development environment that closely models *real world* networks. This requires scalability to the order of hundreds of thousands of virtual network nodes, where a network node is an end-host, a router or a network switching device.
2. To support execution of unmodified protocol and application code.
3. To integrate direct code execution and protocol simulation within a single framework. In the traditional model, direct code execution operates in real time and hence lacks the controllability of virtual time simulation.

The above goals present several interesting research challenges. First, given the scalability goal - support for hundreds of thousands of virtual nodes - we need to analyze its impact on the architecture of the ONE. Even on a cluster computer with hundreds of physical nodes, we need the ability to emulate hundreds of virtual nodes – each with multiple applications and a protocol stack - on a single physical CPU. At this level of scalability, process context switching time becomes a major bottleneck.

The second major challenge stems from the need to support the direct execution of unmodified application and protocol code *and* network simulation. A direct code execution environment (**DCEE)** eliminates the verification and validation problems associated with converting the event driven simulated version of the protocol to a real-world code implementation. However, this does not mean that DCEEs solve all problems, and network simulation is no longer necessary. Distributed network protocols, such as routing protocols are usually modeled as state machines. This makes it easier to convert the model into the event driven form needed for network simulation. An iterative design process is then used to refine the protocol and finally it is implemented in code. It is easy to see why a DCEE will be a cumbersome hindrance to this process. At each stage, the state machine has to be converted to code form before it can be tested - an unnecessary burden. The goal of the ONE is to present an integrated environment that supports both direct code execution of network protocols as well as event-based simulation.

Support for the above goals argues for a component-based design. However, component technologies such as CORBA and COM/DCOM require significant *buy-in*, in the form of commitment to a particular style of programming or an implementation technology. This is a serious hurdle for large, legacy code bases, which cannot be easily ported to a new modeling framework. The primary requirement here is for a *transparent* framework that supports modeling, composition, execution, adaptation, and control of high performance codes, which leads us to the Weaves reconfigurable programming framework.

## 4   The Weaves Reconfigurable Programming Framework

The design goal of the ONE requires support for multiple virtual hosts – each with a protocol stack and multiple applications - within a single process. To operate within the confines of the direct code execution paradigm, the above design goal necessitates a new programming paradigm. We will illustrate this need by a series of examples culminating in a paradigm that can support the compositional model shown in Figure 2.

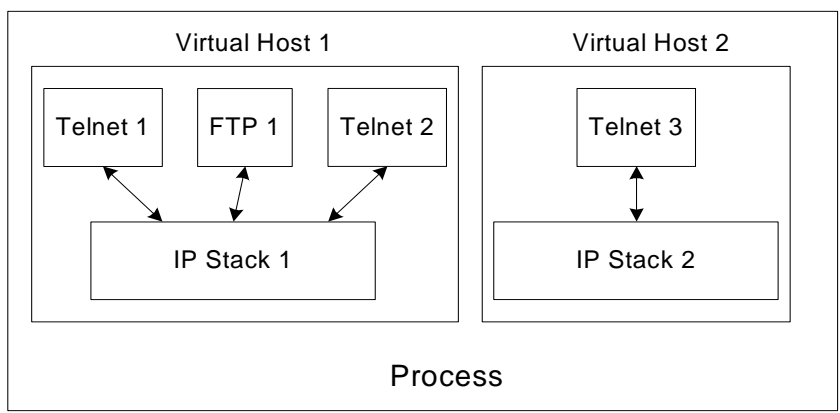

**Figure 1:** Design goal of the Weaves framework. We need to support multiple unmodified network applications linked to protocol stacks, within a single process.

Before we attempt to design a programming model that can support the interaction shown in Figure 2, let us start with a case depicted in Figure 3. Here, two telnet applications are linked against a single IP stack, emulation one virtual end-host.

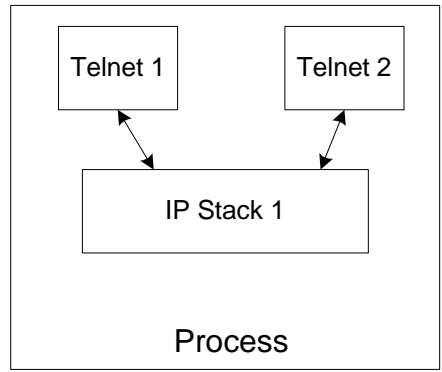

**Figure 2:** A simple compositional model with 2 telnet applications linked with a single IP stack. The composition creates a single virtual host.

In order to model the composition shown in Figure 3, let us start with a simple case, which we call *process per virtual node* model as shown in Figure 4a. This model allows a single network application – say telnet – to be linked to an IP stack. The external references of the telnet application will be bound to the stack, which then uses an underlying PDES layer. Multiple such processes can be run on a workstation to create a virtual network. This simple approach has several problems. First, we cannot link multiple telnet applications to the same IP stack without modifying the application. Since the IP stack in a network node is shared between all applications running at that node, traffic from the various applications can interfere with each other. For instance take a network node that supports both real time applications such as videoconferencing along with best effort applications such as FTP. The traffic from the real time application interferes with the best effort FTP application within the IP stack, resulting in possible less than adequate performance for the real-time application. This effect cannot be studied in the simple process per virtual node model presented here.

The second major problem with the process per virtual node model arises from scalability concerns. A large virtual network will involve several tens to hundreds of virtual nodes, each of which is a process. Inter-process context switch time will be a major concern.

Finally, the PDES algorithm needs to synchronize virtual time between all the processes that constitute the virtual network. If we use a process per virtual node, we increase the number of instantiations of the PDES algorithm and hence the number of independent time lines. This increases the chances of virtual nodes going out of synchronization, causing temporal rollbacks. If we can support multiple virtual nodes within a process, all virtual nodes within the process use a common timeline and hence do not cause unnecessary rollbacks. This reduces the scope and scale of temporal rollback to distinct physical processors participating in the simulation.

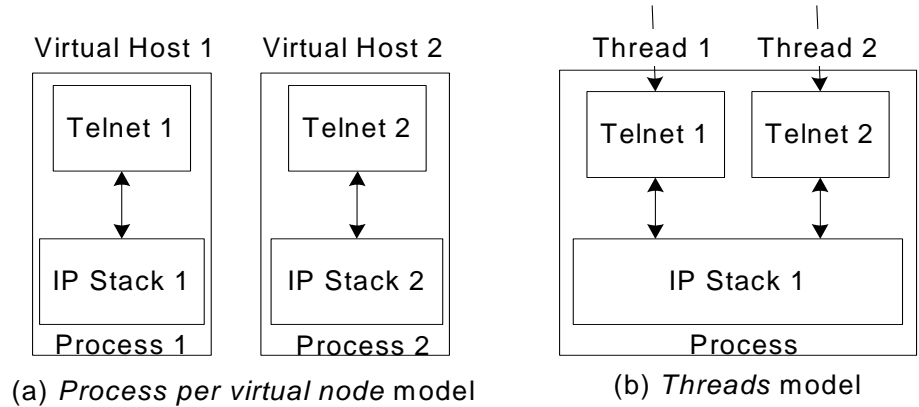

(a) *Process per virtual node* model  (b) *Threads* model

**Figure 3:** The composition shown in figure 3 modeled under (a) process per virtual node model and (b) threads model. Neither of these models can achieve the composition shown in figure 3.

To address the above concerns, let us build a new model based on threads as shown in Figure 4b. In the threads model, the composition shown in figure 3 can be achieved by modifying the telnet application source to create two threads, or inserting a piece of stub code that creates two threads, with the start function of each thread set to the entry point - main() - of the telnet application. As before, the emulated IP stack linked to the application will resolve external references to the IP layer.

The major problem with this approach arises from updates to global variables, in particular telnet is not thread safe. Since threads share global variables, a telnet thread modifying a global variable will inadvertently change the state of the other – unrelated - telnet thread causing erroneous behavior. Ideally, we need 2 copies of all the global variables used in the telnet application. In programs that are explicitly threaded in design, sharing of global state in intentional. In our case, this *sharing is neither intentional nor necessarily desirable*. On the other hand, to create a shared IP stack, we need to share global state within the IP stack between the threads running through the IP stack. We also need to serialize access to global state through the use of mutex locks, which involves significant modifications to the source code. Even if we could make these modifications, we cannot ensure that the resultant code is deadlock free. In effect, we are back to our original verification and validation problem faced in network simulation. There is no easy way to formally ensure that the threaded version of the network application is equivalent to the unthreaded version.

The threads example illustrates the crux of our problem – *conflicting needs* – the need to **avoid sharing** *global state between threads of the telnet application and the need* **for sharing** *global state between the threads running through the IP stack*. The intuition here is that we need a programming model that allows arbitrary sharing of global state. Such a model can subsume both the thread and process models, since it can allow both complete sharing of global state as in threads as well as no sharing as in processes.

The solution to this problem through the use of a compiler-based approach evolves from our earlier work on *EtheReal*, a real time switch that provides bandwidth guarantees for switched Ethernet networks. One of the major components of EtheReal is an automatic fault detection and recovery mechanism that operates within the temporal confines of a real-time system. The fault detection and recovery mechanism relies on the fast construction of a distributed spanning tree – an algorithm that is critical for Ethernet routing. While the prototype implementation of *EtheReal* over a 3 switch network showed the feasibility of the approach, the size of the prototype network was too small to verify scalability. Secondly, we needed a controlled testbed environment to verify the correct operation of the spanning tree algorithm, in particular its ability to withstand the message loss during spanning tree construction.

To test the scalability and correctness of the spanning tree code, we decided to build a simulator that partially implements a direct code execution environment. The simulator simulates a point-to-point switched Ethernet network with multiple virtual Ethernet switches and exposes an interface similar to the internals of the Linux operating system. This allowed us to directly execute the spanning tree code from the *EtheReal* implementation on the simulator, with one modification. All global state variables in the fault recovery code in the simulation are array versions of their counterparts in the implementation. Each virtual switch accesses its copy of the global variables by indexing into the array with its virtual switch identifier. *In effect, the simulator uses a single copy of the spanning tree code and creates virtual instances of the algorithm by switching the state associated with the code.* The principle here is similar to multi-tasking. If we can capture the current state of a state machine, it should be possible to replace the state to create a new instantiation of the machine.

If we take the next step along these lines, we could avoid the task of manually replacing global variables with arrays, by automating the process. This leads us to the first step towards Weaves reconfigurable programming framework.

## *4.1 Defining the Framework.*

The major components of the reconfigurable programming framework are:

1. **Module**: A module is any object file or collection of object files defined by the user. Modules have:

    a. A **data context**, which is the global state of the module scoped within the object files of the module.

    b. A **code context,** which is the code contained within the object files that constitute the module. The code context may have multiple entry point and exit point functions.

2. **Bead**: A bead is an instantiation of a module. Multiple instantiations of a module have independent data contexts, but share the same code context.

3. **Weave**: A weave is a collection of data contexts belonging to beads of different modules. The definition of a weave forms the core of the reconfigurable programming framework. Traditionally, a process has a single name space mapped to a single address space. Weaves allow users to define multiple namespaces within a single address space, with user-defined control over the creation of a namespace.

4. **String**: A string is a thread of execution that operates within a single weave. Similar to the threads model, multiple strings may execute within a single weave. However, a single string cannot operate under multiple weaves. Intuitively, a string operates within a single namespace. Allowing a string to operate under multiple namespaces would violate the single valued nature of variables.

5. **Tapestry**: A **tapestry** is a set of weaves, which describes the structure of the composed application. The physical manifestation of a tapestry is a single process.

The above definitions have equivalents in object-oriented programming. A module is similar to a class and a bead - which is an instantiation of a module – is similar to an object. Tapestries are somewhat similar to object hierarchies. The major exception is that interaction between beads within a tapestry involves runtime binding. We chose to use our own terminology to (a) avoid overloading the semantics of well-known OOP terms and (b) avert the implication that the framework requires the use of an OOP language.

Strings are similar to threads in that (a) they can be dynamically instantiated and (b) they share the same copy of code. However, unlike threads, strings do not share global state. Each string has its own copy of global state. The main goal here is to avoid inadvertent sharing of state between unrelated instantiations of an algorithm, without having to modify the algorithm.

Since strings are an intra-process mechanism, we will illustrate their operation by comparing and contrasting them to threads. A thread's state consists of (a) an instruction pointer, (b) a stack pointer and (c) copy of CPU registers. Each thread within a process has its own stack frame that maintains local variables and a series of activation records that describes the execution path traversed by the thread. When a thread is created, the thread library creates a new stack frame and starts execution at the first instruction of the function specified by the thread instantiation call. When the thread scheduler needs to switch between threads, it saves the current IP, current stack frame and the values in the CPU registers, switches to the state of the next thread and starts execution at the IP contained in the thread state.

Strings involve an extension[1] to the operation of threads. Similar to threads, each string has its own stack frame, which maintains local state. In addition each string also has a copy of the global variables in an area called the ***weave context frame***, the start of which is pointed to by a ***weave context frame pointer***. A weave context defines the namespace for the string. This includes the global variables of all the beads traversed by a string. Note that some of the beads in a string may be shared between strings.

A string's state consists of (a) an instruction pointer, (b) a stack frame pointer, (c) copy of CPU registers and (d) a weave context frame pointer. When a string is created, the string creation call creates a stack frame and a weave context frame (if necessary) and copies the current state of the global variables into the weave context frame. The string creation call also associates a numerical identifier with the newly created string. Since creating a string involves copying its global variables, the string creation cost depends on the storage size of the global variables resulting in a higher creation cost than threads. We justify this cost by noting that this is a one time cost paid at program startup. Also, well-written applications are generally frugal in their use of globals, which mitigates the impact of the copy operation.

Similar to the thread scheduler the string scheduler starts execution of the new string at the first instruction of the function specified by the string instantiation call. When the string library needs to switch between strings, it saves the current IP, current stack frame pointer, the values in the CPU registers, and the current weave context frame pointer, switches to the state of the next string and starts execution at the IP contained in the string state. The inter-string context switch cost is identical to threads. Selective sharing of state in our framework operates at the level of individual beads. We illustrate the operation of selective sharing with the example shown in figure 2 (also repeated in the figure 5 below). The tapestry defines 4 weaves <Telnet 1, IP 1>, <Telnet2, IP 1>, <FTP 1, IP 1> and <Telnet 3, IP 2>, and 4 strings, with each string operating within a single weave. Note that the data context of bead IP 1 is shared across 3 distinct weaves. At run time, context switching between the strings automatically switches the namespace associated with the string, preserving the sharing specified in the tapestry.

---

[1] We will qualify this statement shortly.

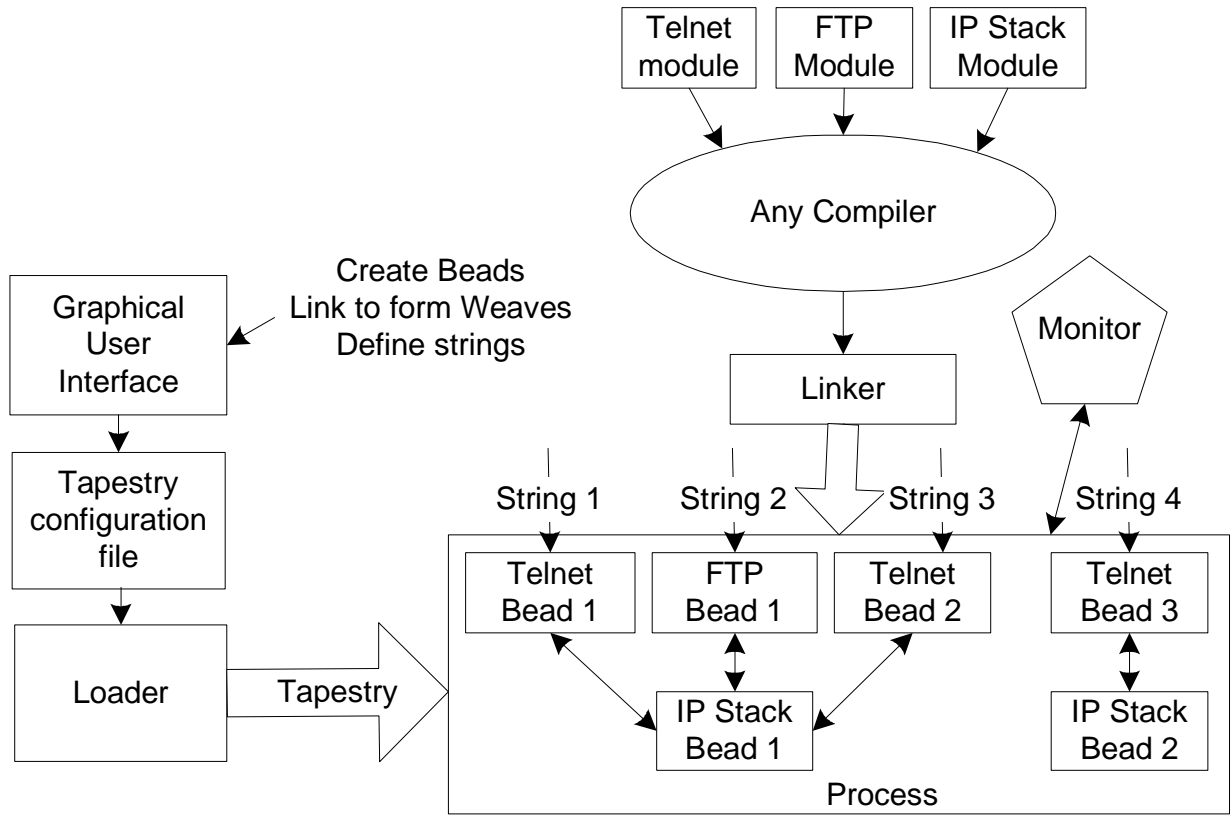

**Figure 4:** Interaction between the various components of the Weave framework.

Figure 5 depicts the design process in the weaves framework. The design process involves two entities: a *programmer* who implements the modules and a *composer*, who uses a graphical user interface to instantiate beads and define the various weaves and strings. The result of the GUI composition is a tapestry configuration file, which is used to load and execute the composed application. Each composed application also has a module called a *monitor* that is automatically linked with the composed application. In the process model, utilities like *ps* (in UNIX) can be used to query the run time of the process. The monitor provides a much more powerful IPC (Inter Process Communication) interface to such functionality. Utilities can query the monitor to determine the current tapestry, beads, strings and weaves within a composed application.

The tapestry generated by the GUI is by not necessarily a static composition. The reconfigurable programming framework allows applications to rewire themselves on the fly in response to dynamic conditions. Two forms of dynamic application composition are supported in the framework. In the first form, if the requisite modules are already linked into the original tapestry, weave-aware applications can modify their structure by creating new beads, defining weaves and instantiating strings at run-time. For non-weave aware applications, the interface exposed by the monitor can be used to modify the tapestry of a composed application. These modifications may be manually made by a user at the command line or can be automatically generated by an external *resource monitoring* agent.

In the second form of dynamic composition, new code modules can be inserted into a running application through a modified dynamic library interface. In this mode of operation, the dynamically inserted code is analyzed at run-time. Dynamically inserted modules can be used in the same manner as statically inserted modules. This interface provides the full capabilities of weaves, including arbitrary namespaces and compositional capabilities, in a run-time compositional framework.

The astute reader may have observed that all the examples of composition presented thus far have an *inverted tree* structure, with sharing occurring at the lower layers. Intuitively, the inverted tree structure appears "natural." This is because it mirrors the sharing and flow of control relationships between processes in an operating system running on a single processor architecture. The resources within the operating system are at a lower layer and shared between the independent processes, which naturally leads to an inverted tree structure.

While the Weaves framework is capable of implementing regular tree structures and context switching between the various strings, from a usability perspective, scheduling in a regular tree structure requires application support. In an inverted tree structure, string scheduling decisions can be made independently of the application without impacting either usability or correctness. However, in a regular tree structure the same does not hold true. Consider this problem from the perspective of the application programmer. A call to a function in a regular tree structure would map to two or more independent data contexts. Choice of the correct data context can only be made by the programmer. The solution in the Weaves framework is to provide support functions that allow the programmer to (a) query the string for its identifier and (b) base the path of execution on the string identifier. This solution is similar to the SPMD programming model on distributed memory architectures, where the execution path is determined by the processor identifier.

Another issue in the Weaves framework arises from reentrancy concerns at the lower layers of an inverted tree. In particular, the problem arises when preemptive scheduling strategies cause reentrancy in beads shared at lower layers, when the codes are not reentrant. We solve this problem by organizing strings into equivalence classes, where each equivalence class contains strings that share beads. Preemptive scheduling switches between strings of different equivalence classes. If the preempted string has not traversed a shared bead, preemptive scheduling can also switch between strings of the same equivalence class.

In our problem domain of the ONE, the virtual network being emulated corresponds to a particular tapestry of weaves. The tapestry is specified by a configuration file that lists (a) the modules, (b) the number of instantiations of each module – with a unique numerical identifier assigned to each instantiation (bead identifier) (c) the namespaces (weaves) which define the relationship between the beads - identified by their weave identifiers and (d) the strings operating under various weaves, identified by their string identifiers. The tapestry configuration file is used by the ONE loader to:
- load the necessary modules,
- instantiate the appropriate number of beads
- setup the tapestry describing the relationship between the beads
- start execution at the entry points of the top level strings.

The tapestry configuration file generated by the graphical user interface allows users to setup complex virtual networks. Since the ONE is based on the fundamentally modular Weaves framework, it is easy to conceptually map the functionality of the GUI to the tapestry configuration file. The framework's support for direct code execution enables us to exploit the vast repository of open source implementations of network applications and protocols to setup a fully functional virtual network environment.

## 4.2 Tuple Spaces

The notion of selective state sharing in the reconfigurable programming framework presents a very powerful mechanism for defining namespaces. Since the definition of a weave permits any set of beads to define a namespace, any composition that can be represented by a connected graph (or a set of independent graphs) can be realized by this framework. From an application's perspective, the definition and operation of distinct namespaces is transparent. This mechanism presents a powerful compositional framework for any procedural code.

The reconfigurable programming framework also supports the notion of shared tuple spaces. In the current definitions, distinct beads of the same module have different data contexts, i.e. data sharing occurs at the granularity of an entire module. To create a shared tuple space, we need fine grain control over the individual members of a data context.

In order to support shared tuple spaces, from the perspective of the framework, we need mechanisms to (a) define a shared tuple space and (b) to selectively share the members of the tuple space across multiple beads. To define a shared tuple space, application composers can use a graphical user interface to denote the members of the tuple space or code modules can use a syntactic notation to mark the members of the tuple space. This information is used at bead creation to merge references to shared members of a tuple space.

### 4.3 Automatic Checkpointing and Recovery

The primary goal of the reconfigurable programming framework is to support adaptive applications that can rewire themselves dynamically in response to changing conditions. Our view of adaptivity encompasses greedy optimistic algorithms that try to take the best execution path given a set of available options. However, the path chosen may not always be right, requiring applications to rollback to a known correct state.

Traditionally, state checkpointing and restoration has been left to individual applications. This significantly adds to the complexity and maintainability of such code bases. Furthermore, event driven codes add an additional layer of complexity. Since the path of execution through an event driven application is not known apriori, checkpointing and restoring such applications present significant challenges. Our goal here is to provide a *transparent support* framework that can checkpoint and recover state, *without application support*.

To provide support for automatic checkpointing and recovery, note that in the weaves framework, each string maintains its global variables in the weave context frame and local variables and call invocation history in the stack frame. This compartmentalizes static state into two well defined regions. We can save the contents of the stack and weave context frames, effectively saving static state. However, this does not still account for dynamic memory allocated during runtime.

To track dynamic memory allocation, we use a mechanism similar to the one used by memory leak debuggers. We overload the library calls responsible for dynamic memory allocation – malloc(), calloc(), realloc() and free() in C. The overloaded calls keep track of the bead identifier, the start of the memory region and the size of allocated memory.

We now have access to both the static as well as dynamic state of the tapestry, which can be used to implement checkpointing and recovery. The naïve mechanism for checkpointing involves (a) saving the contents of the stack frame, (b) saving the contents of the weave context frame and (c) copying the contents of all dynamically allocated memory regions. Restoring application state involves (a) garbage collection of all dynamic memory allocated after the checkpoint and (b) restoring state saved during the checkpoint. It is easy to see that the naïve approach is not memory efficient, particularly in our domain where tapestries can contain hundreds of beads.

A more sophisticated approach to checkpointing involves the creation of a new system call. The observation here is that operating systems already have efficient mechanisms for handling process fork calls, through the use of *copy-on-write* semantics. Our approach is to create a new light weight version of the copy-on-write mechanism that operates in an intra-process domain. When a checkpoint is invoked, the new system call will mark all data pages corresponding to dynamically allocated memory and mark the weave context frame read-only. As the application proceeds, updates to read-only data will cause page faults, which are handling by duplicating the offending page and allowing read/write operation on the duplicate. This mechanism works optimistically, limiting the memory overhead of checkpointing to only modified data.

## 4.4 Run-Time String Migration

Reconfigurable programming is inherently a parallel run-time compositional framework. The examples above show the operation of the framework within the context of a single process. In this mode of operation, the framework provides load balancing on shared memory multiprocessor architectures by executing different strings on different processors. This form of load balancing is transparent to the design of the application.

Since the current generation of large parallel supercomputers is based on distributed memory design, we need to extend the framework to provide similar *transparent* load balancing capabilities to distributed memory codes. Currently, distributed memory codes implement their own load balancing. Since we know the structure of the composed application, we are in a position to both expose additional interfaces as well as augment existing load balancing capabilities.

In large tapestries composed of tens of thousands of beads, good load balancing is necessary to obtain reasonable speedup. The scale of the system combined with often, incomplete knowledge of runtime load makes it nearly impossible to statically load balance such a system. Dynamic load balancing guided by runtime analysis of load is necessary to ensure scalability. Our view here is that load balancing in a distributed memory environment translates to run-time code migration. This will ensure that all participating host processors see equal computation and communication loads.
Our target environment – the ONE – runs unmodified network applications/protocols, which precludes the use of data distribution strategies for load balancing. The main strategy at our disposal is code migration. Our support for direct code execution also means we can't rely on application support for code migration. The main issue here is determining the resolution of code migration. Should code migration occur at the level of individual beads or something larger?

To answer this question, let us take a look at the issues involved in code migration at different resolutions. In the case of an individual bead, we can track its static state but it is much harder to keep track of its dynamic memory allocation. To see why, let's take the example of a bead that invokes a function within another bead. The target function allocates an array of pointers, allocates memory to each element of the array and returns a pointer to the start of the pointer hierarchy. The overloaded memory allocation calls from section 2.1.3 will incorrectly attribute the memory allocated to each element of the array to the bead corresponding to the target function. Without exhaustive analysis and significant runtime support, it is not possible to track dynamic memory allocation within a bead.

To avoid the above problem, our observation is that while it may not be possible to track the dynamic memory usage of a single bead, it may be easier to track the total dynamic memory used by a set of beads. In essence, we are trying to create closed regions of interacting beads – *an island of beads* - that exchange memory between them, but have no connection to other beads. If we apply this idea to our problem, we can immediately see the island of beads – all beads within a single virtual host. Since virtual hosts in the real world are physically separate entities, they cannot share memory and this is true in our emulated world as well.

Although the above solution of tracking dynamic memory at the resolution of virtual hosts rather individual beads is specific to our problem, we believe that in general such islands of interacting beads can be found in most applications – they represent entities at a higher layer of abstraction. Graph theoretically, an island of beads represents a closed graph with no external connections. The modular framework of weaves also aids the process of isolating islands of beads by forcing designers to look at beads interactions during application composition. The GUI front-end used to create a tapestry can also be used to mark specific islands of weaves, which then become targets for code migration.

Our target platform – a workstation cluster – uses the SPMD model of execution. Since the same program executes on all cluster nodes, the necessary code modules either already exist at the target of the migration, or can be instantiated at run-time through the dynamic library interface. Code migration then reduces to the problem of instantiating new weaves at the target node corresponding to newly migrated island and migrating the state corresponding to the island. Since weaved code always uses indexed addressing, the migrated code does not need any code patching to be functional.

The above analysis ignores a very serious problem with intra-process code migration – *pointer aliasing*. Traditionally code migration has been handled at the process level resulting in process migration. Since each process operates within its own address space, when a process is migrated, it sees the same virtual memory addresses on the target processor. In our domain, migration occurs at the level of string, which is an intra process entity. When a string is migrated, it will not necessarily get the same virtual memory addresses on the target processor. Code patching will be needed to fix the addresses. This significantly complicates migration in distributed memory machines.

This problem is even more complex than the above description. To see why, let us take a case where a bead allocates dynamic memory to a pointer variable *ptr*. It then sets a second pointer variable *ptr1* to *ptr*, i.e. *ptr1* points to the same memory location as *ptr*. When we migrate this code, we allocate dynamic memory at the target machine and set *ptr* to point to this memory. However, *ptr1* is still pointing to the old memory location from the source processor. There is no guarantee that the same memory location in *virtual memory address region* is available on the target processor. Not only do we have to fix memory addresses allocated dynamically, we also need to ensure that all aliases of memory addresses are fixed appropriately, a problem known as pointer aliasing.

Pointer aliasing is a significant research issue. Current solutions are based on restricting source language semantics to prevent pointer aliasing, or executing code within virtual machines. Neither of these solutions is available to us. We have no control over the source language and emulating code over a virtual machine will impose unacceptable performance penalties.

To solve this problem, we propose a solution called *shared virtual memory*. The observation here is that pointer aliasing becomes an issue because of the shared nature of the virtual address space from the perspective of intra-process migration. In our solution, we statically allocate regions of the virtual address space to participating processors. The first processor allocates dynamic memory in the region 1 – X MB, the second processor allocates memory in the region X MB – 2X MB and so on, where the memory addresses are in virtual memory space. In this model, when a string migrates from a source processor to a destination processor, it is guaranteed that memory addresses in its VM space are available on the target processor. This solution effectively bypasses the pointer aliasing problem. [2]

The sharing of single VM address space across multiple processors imposes size restrictions on the VM addresses that can be allocated to any single processor, which in turn impacts the dynamic size of an application. However, this is not as restrictive as it appears at first sight. On a 64 bit processor, we can allocate 1 TB of VM space to each processor and still support parallel applications that can run on 16 million CPUs – well beyond the scope of current applications and super computers (the calculation divides the 64 bit space into 40 bit VM addresses and 24 bit CPU identifiers). As the dynamic memory demands grow, we can allocate more bits to the address space, reduce the maximum number of CPUs that can participate in the computation and still stay ahead of Moore's Law.

---

[2] We also looked at dynamic partitioning schemes. The problem is that it increases the cost of dynamic memory allocation calls. Hence we propose a static partitioning scheme.

Performing a similar calculation for 32 bit processors shows that the above scheme imposes significant restrictions. For instance, if we allocate 1 GB of VM space per processor, we can only support applications that can run on 4 processors, which is definitely not acceptable. To get around this issue, we note that the main 32 bit processor family is based on the Intel x86 instruction set. Starting with Pentium Pro™ family, Intel™ added 4 additional bits to the addressing, resulting in a 36 bit VM address space. This allows us to allocate the full 32 bit address space to each processor and support parallel applications up to 16 CPUs.

Even with the additional 4 bits of VM addressing space, limiting parallel applications to 16 CPUs is overly restrictive. To ameliorate this condition, we compartmentalize CPUs into *VM regions* of 16 CPUs each. Strings can freely migrate within a region and with some restrictions, even across regions. This solution offers an attractive trade-off between scalability and run-time load-balancing for distributed memory architectures.

# 5   Implementation and Evaluation

The core of the Weaves DCE framework is the abstraction of a weave, which allows an application composer to define arbitrary namespaces over a composed application. To implement the weave abstraction, we need a data structure that can efficiently capture the state separation and state recombination needs of the DCE framework.

Before we discuss the specifics, note that the goals of the DCE framework place additional constraints on the implementation of the weave abstraction. First, our transparency requirement states that the solution should be transparent to the application. Since, the application may be written in any programming language, the transparency requirement precludes modification to the source code to implement the namespace abstraction. Second, from a scalability perspective, the implementation should be efficient. In particular, we need to minimize context switch time between the various namespaces defined in the composed application.

To meet the transparency requirement, the implementation of the namespace abstraction works by analyzing the Executable and Linking Format (**ELF**) object files produced by any compiler. ELF is a public domain file format used to represent both object code as well as the final executable on most UNIX systems. Our current prototype is implemented on the Linux operating system running on Intel x86 architectures. Since the implementation only depends on the ELF file format, it can be easily ported to other operating systems/architectures. Furthermore, we anecdotally note that the features of the ELF file format used by our implementation are common to object file formats. Hence, it should be possible to extend the prototype to support other object file formats as well.

The ELF file format uses the Global Offset Table (**GOT**) data structure to access global state in an application. The GOT data structure maintains an array of *pointers* (instead of data values), with each pointer referring to a global data variable. To access data, applications first index into the GOT data structure to get a pointer to the data and then use the pointer (and possibly an offset) to retrieve the data value. The number of entries in the GOT structure is proportional to the number of variables and is independent of the size of each variable. For instance, an array variable has a single entry in the GOT structure. The observation here is that the GOT defines the namespace of the application. Typically, a program contains a single GOT structure reflecting the single namespace within an application. However, by appropriately defining multiple GOT structures, it should be possible to create multiple namespaces within a single ELF executable.

The problem with the basic GOT structure is that compilers hardcode the base address of the GOT structure and the index into the GOT at compile time. To implement multiple namespaces, we need to create multiple GOT structures and, at runtime, copy them over to the fixed base address generated by the compiler. This operation is expensive since its cost is proportional to the number of global variables, which can potentially be large.

Instead, we note that compilers produce relocatable code (for instance, the command line option –fPIC on the gcc family of compilers) to support dynamic libraries. In relocatable codes, the base of the GOT structure is pointed to by a base register. All indexed accesses into the GOT are made with reference to the current value of the base register. The use of relocatable object code and indexed access to the GOT forms the basis of our implementation.

To implement the weave abstraction, we create a new GOT structure for each distinct weave in the composed application. To implement state separation between beads belonging to different weaves, we first create copies of the data and point the GOT entries in the weaves to the distinct copies of the data. To enable state recombination between weaves sharing a bead, we set the pointer in the GOT entries in the different weaves to point to the same data value. The double indexed nature of the GOT structure enables state separation/recombination at the resolution of a single data variable, which can be used to implement arbitrary data sharing at both the tuple space and module levels.

To implement the string abstraction, note that a string is really a thread operating under a user specified namespace. Since we have a mechanism to create the namespace, context switching between strings involves context switching the thread state and switching the namespace. What we need here is an efficient mechanism for switching namespaces.

To switch namespaces, we note that the GOT structure is accessed through a base register (%ebx in our current implementation). Hence, context switching between namespaces merely involves changing the base register to point to a different GOT structure, a single instruction move operation. Furthermore, we note that we do not even have to pay the cost of this instruction. The base register is automatically saved and restored by the thread context switch routine. If we initialize the base register to point to different GOT structures before thread creation, the thread context switching routines automatically context switch between weaves. This results in a weave context switch time that is identical to thread context switch time. Our current implementation of the Weaves framework works over both POSIX Threads (pthreads) as well as the GNU Portable Threads (Pth) thread libraries.

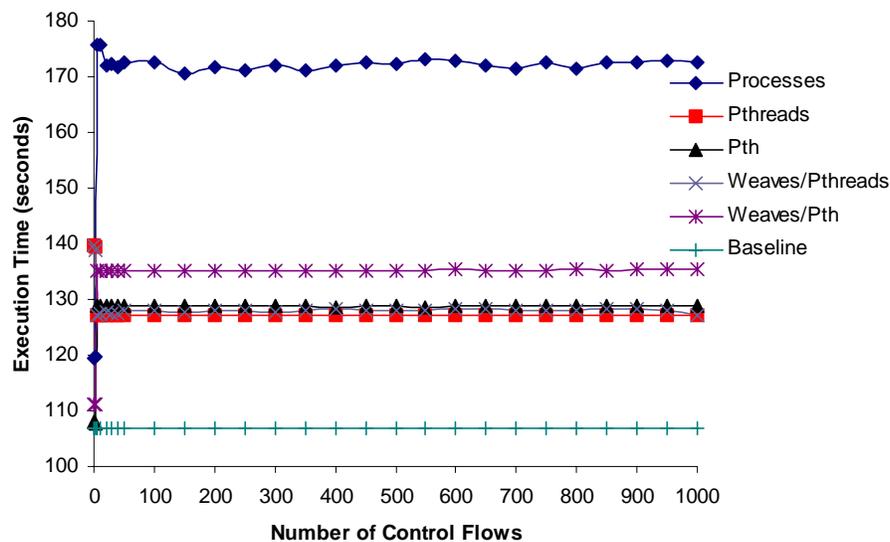

**Figure 5:** Comparison of inter-flow context switch time in the threads, processes, and weaves programming models. The baseline single process application implements a calibrated delay loop of 107 seconds.

We ran a series of experiments to compare the context switch time under the threads, processes and weaves programming models. In this experiment, we created a baseline application that implements a calibrated delay loop (busy wait). We then implemented threads-, processes-, and weaves- versions of the application. In each of these versions, there are $n$ independent flows of control over the same code, where each flow of control executes a calibrated delay loop, which does $1/n^{th}$ the work of the baseline application. We then measure the total time taken to execute the application under each of these models. Since each of the control flows does $1/n^{th}$ of the work and there are $n$ flows, the total time taken should the same as the baseline calibrated delay loop case, except for an additional context switching cost.

Figure 8 shows the results of the experiment on a single processor AMD Athlon™ workstation running the Linux operating system. The results show the run time for five cases: (a) baseline calibrated delay loop, (b) pthreads threads library, (c) Pth threads library, (d) processes, (e) Weaves over pthreads, and (f) Weaves over Pth. The results clearly show that the weaved implementations are significantly faster than processes, even in this simple case, where the copy-on-write semantics of the *fork()* call are very effective. Furthermore, the run time of weaved implementation of pthreads is very close to the base run time of pthreads alone. The marginal variation in runtime is due to the slightly higher weave creation cost, which is included in the run time. Also, the pthreads implementation is relatively efficient, since the Linux kernel includes operating system support for it.

However, in the case of Pth, the run time of the weaved implemented is higher than the base Pth case. This increase in runtime is because unlike pthreads, Pth is a user-level library and hence suffers from timer inaccuracies inherent in user-level library implementation. As an aside, we mention that we noticed several discrepancies in the Linux operating system scheduler. The base processes and pthreads implementations showed far less variation on the SGI IRIX™ operating system.

## 5.1 Weaving Sweep3D

To test the capabilities of the Weaves framework on real-world scientific codes, we weaved the Sweep3D application [Adve et al., 2000]. To support the message passing primitives used by Sweep3D, we created a simple threaded MPI emulator, which implements only the **nine** MPI primitives used by Sweep3D. To ensure correctness, the MPI emulator implementation follows the guidelines set forth in the MPI specification. Our MPI emulator is intended as a test prototype and is neither as comprehensive nor as capable as a complete MPI implementation.

The main characteristic of Sweep3D is that it uses no global variables. Since the application only relies on locals, multiple instantiations of local state should be enough to create a VM abstraction. This characteristic makes Sweep3D inherently thread-safe, which enables it's modeling by either the threads or process models. However, since the application is written in Fortran 77, with dynamic array extensions, modeling with the threads and processes models present interesting implementation problems. While trying to model the application using POSIX threads, we found that there was substantial global state in the .data section of the ELF executable, a compiler issue, which essentially made the code-base "thread unsafe". Weaving the Sweep3D code-base created independent namespaces, resulting in a thread-safe version.

In the weaved implementation, we create $n$ distinct virtual machines, each of which executes an independent instantiation of the Sweep3D application. To do this, we create $n$ distinct Sweep3D beads and $n$ weaves, where each weave has a distinct Sweep3D bead and a shared emulator bead. Each weave also has a single string associated with it. The $n$ distinct virtual machines run on a single processor workstation.

We compared the performance of our single processor weaved implementation of Sweep3D against measured values from real runs for up to 150 processors. Measurements for the real runs were made on our 200 processor cluster *Anantham*. Each node of *Anantham* has a 1GHz AMD Athlon ™ processor, with 1GB RAM. The nodes are interconnected over a switched Myrinet communication fabric, which provides 4Gbps of network bandwidth per node. Since the Sweep3D application performs its own timing measurements, we compared the timing numbers (CPU Time) of the weaved version of Sweep3D with the measurements from actual runs. The two input files (50x50x50 and 150x150x150 decompositions) provided in the Sweep3D distribution were used to drive the Sweep3D application.

For upto 150 processors, the timing results from the weaved implementation and the actual runs were consistent to within 0.2%. Furthermore, we tested the weaved version of Sweep3D with over 1000 weaves on a single processor. The variation in the timing results between multiple runs was within 0.2%. This clearly shows that even at high levels of scalability (over 1000 weaves/processor) context switch time does not impact the efficacy of the Weaves DCE framework.

# 6 Ongoing Work

## 6.1 Weaves for Checkpoint and Run-Time Migration of Parallel Communication Libraries

One of the most challenging cases for automatic checkpointing, recovery, and migration is that of large-scale, distributed-memory, message-passing codes. While this problem is motivated by reliability concerns for parallel codes running on large clusters, it is especially critical for grid computing codes, where checkpointing and run-time migration are essential. In the last two months, we have developed a user-level implementation of the TCP/IP stack (as a part of another project) which, combined with the

Weaves framework, suggests a promising new approach to checkpointing and runtime migration of parallel codes.

There are several serious issues with checkpointing parallel codes written using MPICH (a popular implementation of the MPI library) or PVM. First, we need a reliable single process checkpointing tool; these are not available on all platforms. While a single process checkpointing tool can capture process state, we also need mechanisms to capture state maintained within the operating system on behalf of the process. This includes open network socket handles and network data within operating system buffers. One approach to this problem is to create a consistent global state of the parallel application [1] and then initiate the checkpoint operation. While this approach works, it still doesn't enable migration. MPI implementations maintain significant static environment state within themselves, including IP addresses, hostnames and open TCP connections. To migrate a checkpointed MPI application, we need mechanisms to update the internal MPI state to reflect the change in the underlying environment, which makes any implementation of such a system specific to a particular MPI or PVM codebase. Our goal is to implement a transparent checkpoint and migration framework for any parallel communication library that uses the TCP/IP protocol stack. In our design, a parallel application is linked to a message passing library (e.g., MPI), which in turn is linked to our TCP/IP implementation through a standard sockets interface. Our TCP/IP implementation treats the underlying communication subsystem as an unreliable link and only requires support for two calls, a transmit call and an inbound receive call.

The main advantage of this design is that the user-level TCP/IP stack provides an additional layer of indirection, exposing a common IP address and hostname independent of the physical platform or underlying communication infrastructure. We propose to use the Weaves framework to deliver platform independent checkpointing and migration facilities. The Weaves framework will checkpoint the entire application, including the user-level TCP/IP implementation. When a parallel application is migrated, the user-level TCP/IP implementation is used to abstract the specifics of the target environment from the application. There are several additional advantages to this approach. First, a parallel application can be moved to a platform with a different underlying communication infrastructure. Secondly, the same framework can be used to simulate different physical parallel communication fabrics and analyze the performance of communication libraries. Finally, this approach enables easy portability of parallel communication libraries, since they can be developed for TCP/IP and rely on our system to provide the actual mapping to the physical environment.